\documentclass[aps,prl,nofootinbib,twocolumn,superscriptaddress,letterpaper,amsmath,amssymb]{revtex4}

\usepackage{graphicx}  % needed for figures
\usepackage{dcolumn}   % needed for some tables
\usepackage{bm}        % for math
\usepackage{amssymb}   % for math
\usepackage{feynmf}%%%{feynmp}
\usepackage{slashed}
\usepackage{wasysym}

\def\gsim{\lower0.5ex\hbox{$\:\buildrel >\over\sim\:$}}
\def\lsim{\lower0.5ex\hbox{$\:\buildrel <\over\sim\:$}}

\newcommand{\be}{\begin{equation}}
\newcommand{\ee}{\end{equation}}
\newcommand{\bea}{\begin{eqnarray}}
\newcommand{\eea}{\end{eqnarray}}

\newcommand{\nbox}{{\,\lower0.9pt\vbox{\hrule \hbox{\vrule height 0.2 cm
\hskip 0.2 cm \vrule height 0.2 cm}\hrule}\,}}

\begin{document}

\thispagestyle{empty}
\vspace*{-3.5cm}

\vspace{0.5in}

%\begin{flushright}
%\today\\
%\end{flushright}
%\vspace{0.5in}
\title{Searching for Spurious Solar and Sky Lines in the Fermi-LAT Spectrum}

\begin{center}
\begin{abstract}
We search for a unified instrumental explanation of the spectral
features seen near $E_\gamma=130$ GeV   in photons collected by 
Fermi-LAT from the galactic
center and from the Earth's limb.  We report for the first time a
similar feature in photons originating from the vicinity of the Sun,
and examine the instrumental characteristics of this Solar feature.  To test an instrumental
hypothesis, we identify the range of
photon incident angles where most of the peak photons are observed in
these three spectral features. An examination of the spectrum of
photons from the rest of the sky with this characteristic angular range reveals a hint of a spectral feature near
$E_\gamma=130$ GeV. These results
cast further doubt on the dark-matter-annihilation interpretation of the galactic center peak.
\end{abstract}
\end{center}

\author{Daniel Whiteson}
\affiliation{Department of Physics and Astronomy, University of California, Irvine, CA 92697}
%\pacs{}
\maketitle

\section{Introduction}

The particle nature of dark matter remains a mystery.  One potential
avenue for discovery is via dark matter annihilation into standard
model particles. If the annihilation results in fermions or heavy bosons,
 photons may be produced via hadronization and the decay of $\pi^0$
particles. These photons are expected to be fairly  low-energy
($E_\gamma \le\approx 50$ GeV) and are expected to be
difficult to distinguish from other sources. A clearer feature may appear from annhilation directly into
two-body final states including a photon. Rather than yielding a broad
energy spectrum,  this process would
produce a photon with a well-defined
energy simply related to the mass of the dark matter particle.   Such high-energy gamma rays typically do not scatter in transit
 to the Earth from the regions of high dark-matter density where they
 are produced, making their
energy and direction  useful handles. The Fermi Large Area Telescope
(LAT) has a large field-of-view and excellent energy resolution over a
broad range of energy (20 MeV - 1 TeV)~{\cite{Atwood:2009ez},
  making it a powerful probe for  peaks in the photon spectrum from
  dark matter annihilation\cite{Abdo:2010nc,Fermi:2012}.

Recently, a statistically significant peak has been reported in the Fermi-LAT
photon spectrum near $E_\gamma=130$ GeV\cite{Bringmann:2012vr,Weniger:2012tx} with a source 
close to the galactic center
\cite{Bringmann:2012vr,Weniger:2012tx,Tempel:2012ey,finksu,Rao:2012fh}. The
importance of such a discovery, if interpretted as due to dark matter
annihilation, requires a careful exploration of other  more mundane
explanations, such as potential astrophysical spectral features in the
non-dark-matter background from the  galactic center, or instrumental effects in the Fermi-LAT detector.  Suspicious
features have been reported in the incident angle of the photons from
the galactic center~\cite{fermisplots} and a statistically significant
feature has been reported in photons from the Earth's
limb~\cite{Finkbeiner:2012ez}, where no signal is expected from dark
matter annihilation.

In this paper,  we consider another
luminous source, the Sun, for evidence of spectral features.  In addition, we use the {\sc sPlots}~\cite{splots} algorithm to reconstruct
 -- separately for peak and background photons 
  -- the distributions of instrumental quantities in an attempt to
  identify a common instrumental characteristic unique to the observed
  spectral features. Such a common characteristic would point strongly
  to an instrumental explanation which would be
  independent of the source of the photons. To test such potential instrumental
  explanations, we examine the spectrum of photons with these
  instrumental characteristics from the entire sky for evidence of spurious features.

\section{The Fermi-LAT data sample}

We use the publically available Fermi-LAT  photon data 
through January 3rd 2013,
making  standard quality requirements~\cite{qual,blah:2012kca}. In
addition, we consider several additional requirements for the distinct samples. The
analysis of the
Earth's limb photons requires a zenith angle (measured with respect to the zenith line,
  which passed through the earth and the Fermi spacecraft) between $110^\circ$ and
$114^\circ$ degrees; the analysis of the Solar photons requires a
zenith angle less than $105^\circ$, rocking angle of less than $52^\circ$,
 and angular distance from the
position of the Sun of less than five degrees.  The galactic center
spectrum is drawn from 
a five-degree circle around ($l=0,b=0$) in galactic coordinates,
requiring zenith angle less than $105^\circ$ and rocking angle of less than $52^\circ$.

Other than the reconstructed energy, the photons have other measured
characteristics~\cite{fermidefs} which may give insight into instrumental effects:
\begin{itemize}
\item incident angle $\theta$, measured with respect to the top-face normal of
  the LAT,
\item  azimuth angle $\phi$, measured with respect to the top-face
  normal of
  the LAT, folded as described in Eq. (15) of Ref.~\cite{blah:2012kca}.
\item mission elapsed time, measured relative to January 1, 2001,
\item conversion type (front or back), indicates whether the event
  induced pair production in the front (thin) layers or the back
  (thick) layers of the tracker,
\item the magnetic field in which the LAT is immersed, as
  parameterized by the McIlwain $B$ and $L$
  parameters~\cite{mcilwain}, 
\end{itemize}

\section{Spectra}

For the  observed feature we assume a single line where the pdf $f_{\textrm{line}}(E_\gamma|E_{\textrm{line}})$ is defined according to the
Fermi-LAT energy dispersion tools definition~\cite{fermiedisp} with a
true photon energy of $E_{\textrm{line}}$ (see Fig.3 of Ref~\cite{fermisplots}).
To analyze the energy dependence, we integrate out the dependence on
$\theta$ using the exposure and livetime in the region of interest.

The background pdf is a simple power-law:

\[ f_{\textrm{bg}}(E_\gamma|\beta,\alpha) = \beta\left(
  \frac{E_\gamma}{E_0}\right) ^{-\alpha} \]

\subsection{Earth's Limb Spectrum}

Some fraction of the time, the LAT's field of view includes the limb
of the Earth, where photons are produced in collisions of cosmic rays with
the atmosphere. These are not salient for dark matter searches, given
their terrestrial source, and so are typically discarded.  The
observation of a statistically significant feature near $E_\gamma=130$
GeV in this spectra for photons with $30^\circ<\theta<45^\circ$
(Ref~\cite{finksu} and Fig.~\ref{fig:signif}) where none is expected
from dark matter annhiliation points strongly to an instrumental
explanation of this feature.

\subsection{Solar Spectrum}
In the same spirit, we examine the high-energy Solar spectrum. The
Fermi collaboration studied the low-energy ($E_\gamma<10$ GeV)
spectrum in a
smaller dataset~\cite{Abdo:2011xn}.

 While arguments
have been made that the Sun's gravitational well may trap dark matter
particles~\cite{Peter:2009qj}, and searches have
been performed for evidence of Solar dark matter annhiliation via
neutrinos~\cite{IceCube:2011ae}, the majority of the dark matter is expected to be
gathered near the center of the Sun.  Photons resulting from Solar
dark-matter annhiliation would not escape the Sun without interacting,
and so no signal is expected in the Solar photon spectrum\cite{Sivertsson:2009nx} unless the
Solar dark matter halo is much broader, contrary to current ideas. A feature
in the Solar spectrum would suggest instrumental causes, in the same
manner as the peak in the Earth's limb data.

To isolate Solar photons, we use the position of the Sun relative to
the LAT and calculate the effective position of the Sun in galactic
coordinates at the moment a given photon is received, ($b_{\astrosun},l_{\astrosun}$). The Solar region
is then defined via the angular distance

\[ \cos (\Delta R_{\astrosun}) = \sin(b_\gamma)\sin(b_{\astrosun}) + \cos(b_\gamma)\cos(b_{\astrosun})\cos(l_\gamma-l_{\astrosun}) \]

We consider photons with $\Delta R_{\astrosun} <5^\circ$ outside the region of
the galactic center (see Fig.~\ref{fig:sun}, left) to be Solar
photons.  The Solar photon spectrum is shown in Fig.~\ref{fig:sun}
(right), compared to the spectrum of the inverted region ($\Delta
R_{\astrosun}>5^\circ$, same veto on galactic center photons).  The local
statistical significance of the feature near $E_\gamma=135$ GeV is 3.2$\sigma$ (see Fig.~\ref{fig:signif_sun}) which does not
suffer from a look-elsewhere effect, as the region-of-interest in
$E_\gamma$ of the feature is
determined by the feature in the galactic center spectrum.  This appears to be the
first examination of the high-energy  Solar spectrum ($E_\gamma>10$
GeV) using Fermi-LAT data.  The
same feature appears for different requirements of $\Delta
R_{\astrosun}$, see Fig.~\ref{fig:sun2}; in fact, masking out the
physical extent of the Sun ($\Delta
  R_{\astrosun} \in[1,5]^\circ$) yields nearly the same spectrum,
  suggesting that these photons do not originate from the Sun proper,
  but from its vicinity.

\begin{figure}
\includegraphics[width=3in]{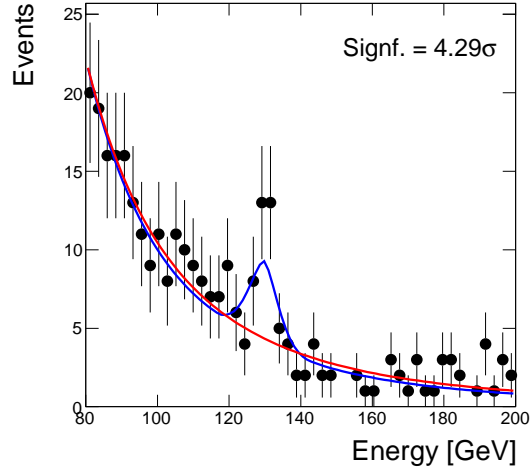}
\caption{Observed Fermi-LAT photon spectrum from the Earth's limb,
  requiring in addition $30^\circ<\theta_{\textrm{photon}}<45^\circ$,
  following Ref.~\cite{finksu}.}
\label{fig:signif}
\end{figure}

\begin{figure}
\includegraphics[width=1.6in]{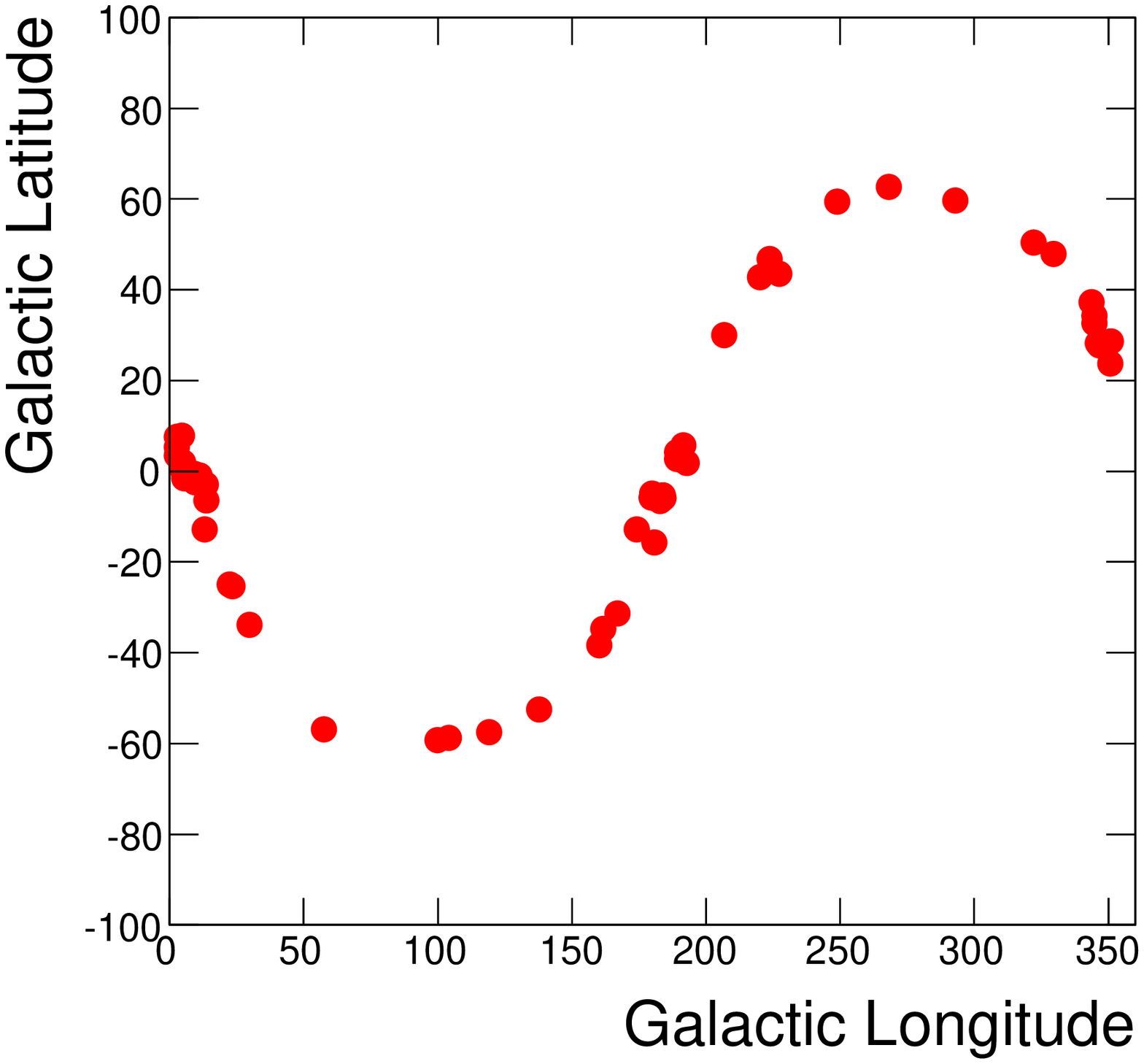}
\includegraphics[width=1.6in]{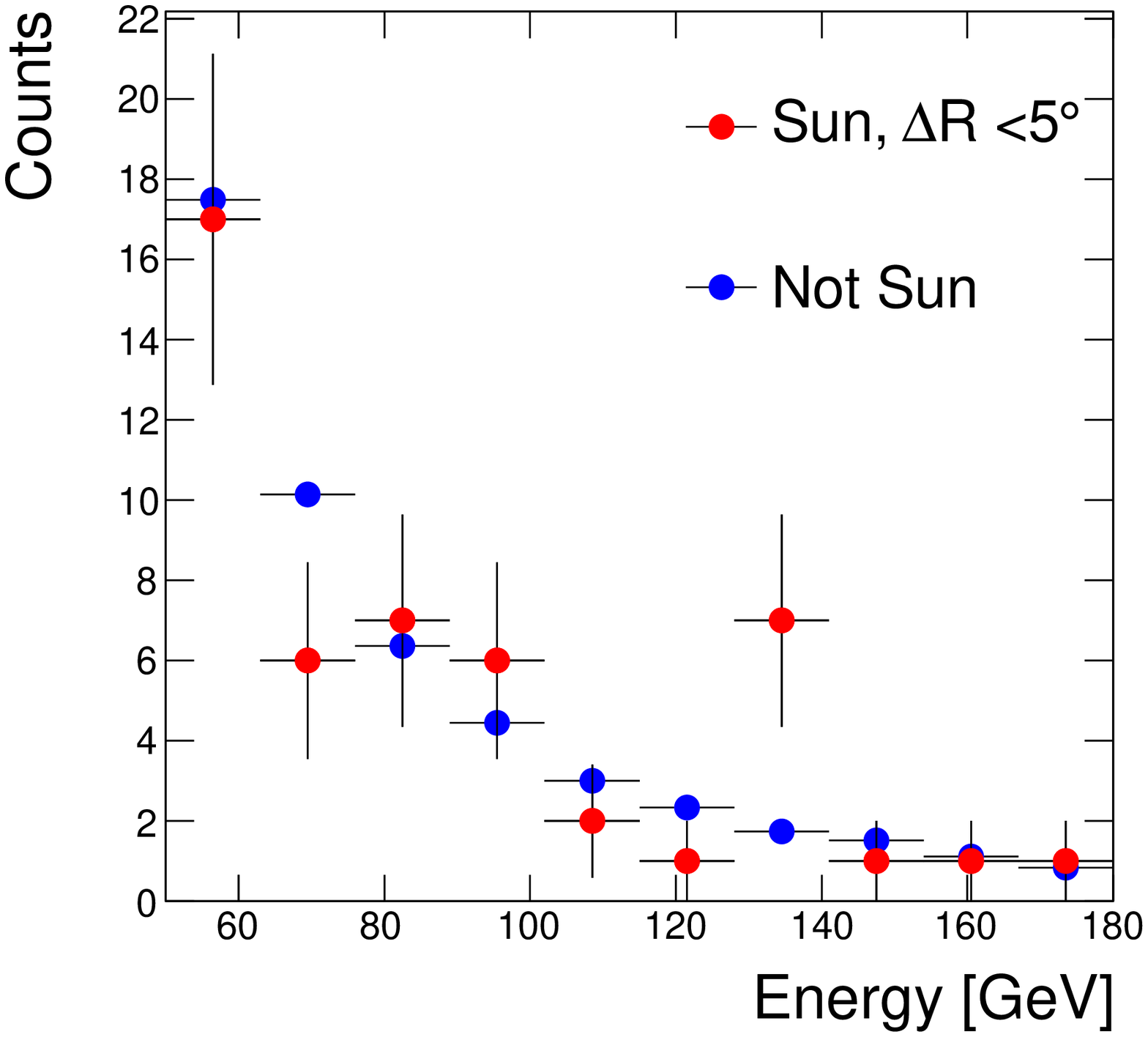}
\caption{Left, galactic coordinates of photons within $\Delta
  R_{\astrosun}<5^\circ$ of the position of the Sun. Right, energy spectrum for
  photons from the vicinity of the Sun ($\Delta
  R_{\astrosun}<5^\circ$, excluding the galactic center),
  compared to photons from the rest of the sky ($\Delta
  R_{\astrosun}>5^\circ$, excluding the galactic
  center), normalized to equal yield.}
\label{fig:sun}
\end{figure}

\begin{figure}
\includegraphics[width=1.6in]{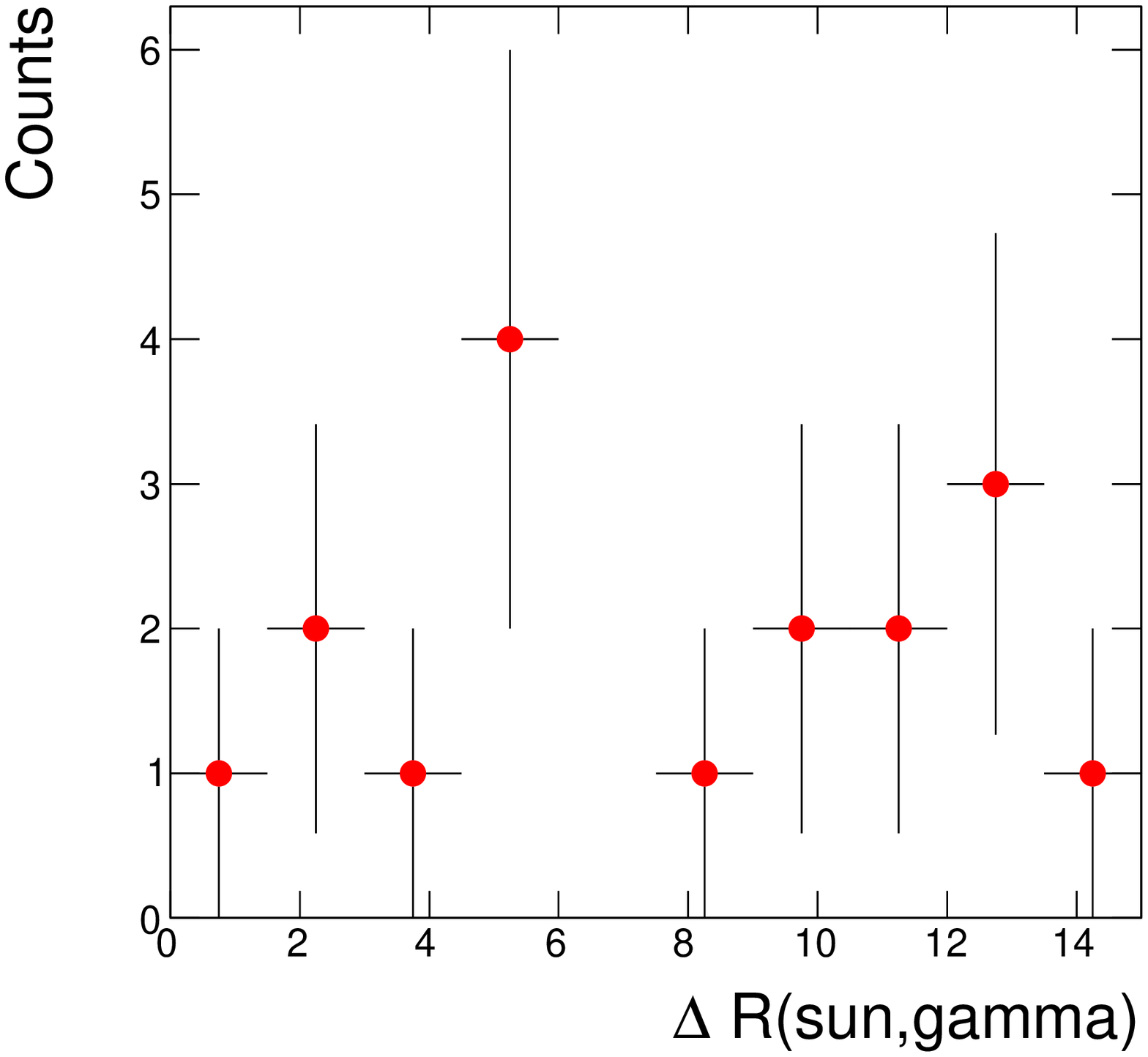}
\includegraphics[width=1.6in]{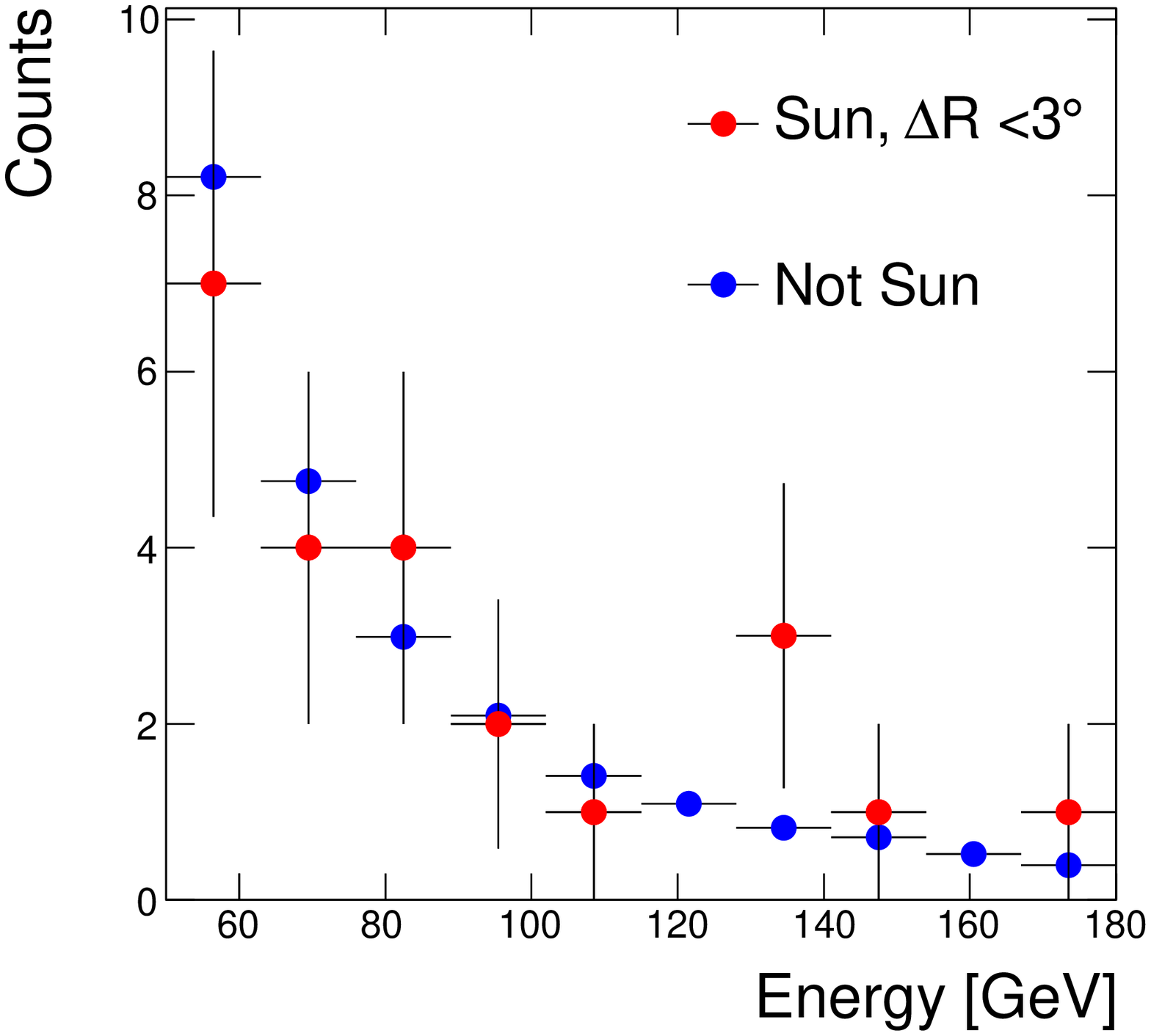}
\includegraphics[width=1.6in]{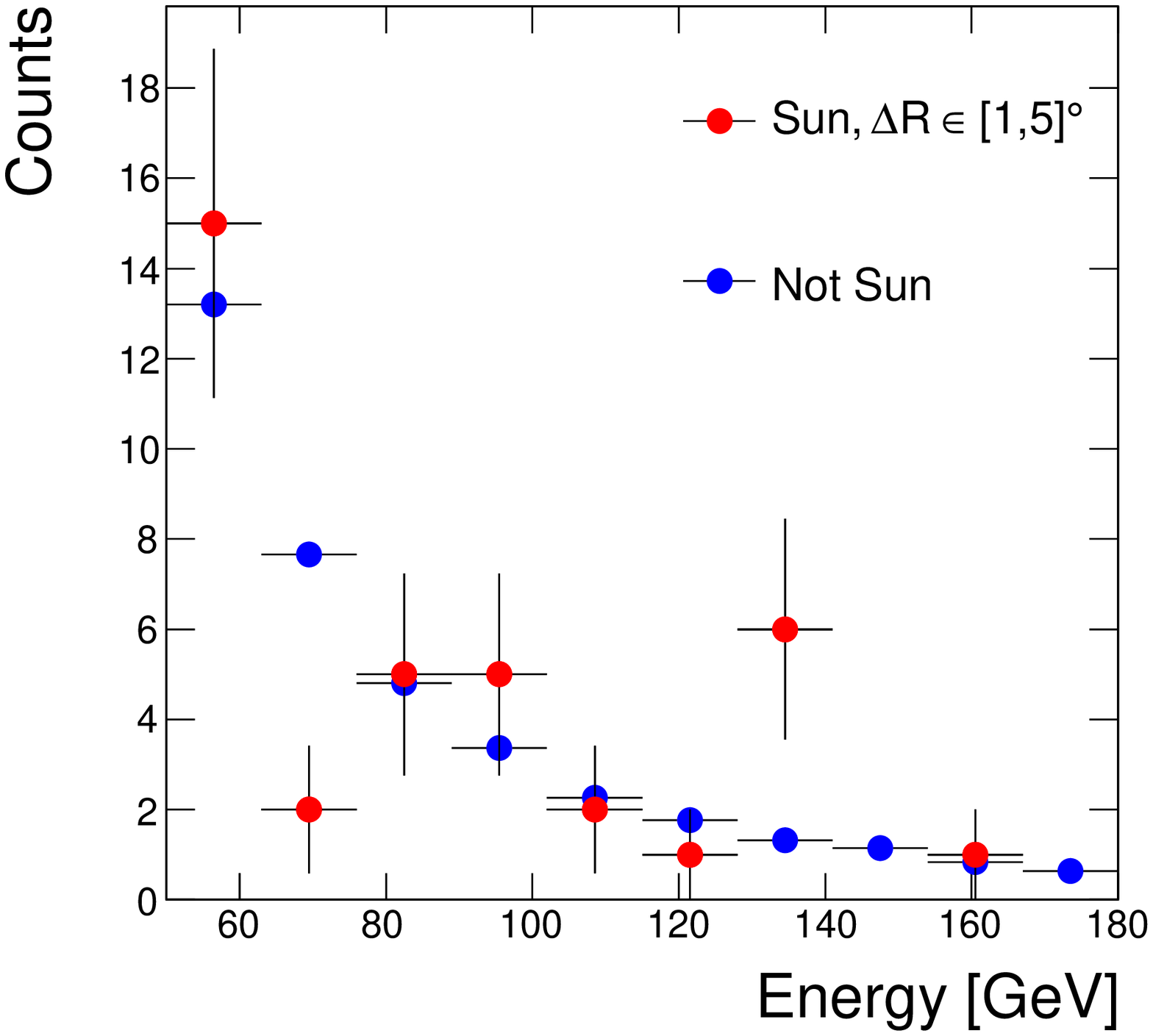}
\includegraphics[width=1.6in]{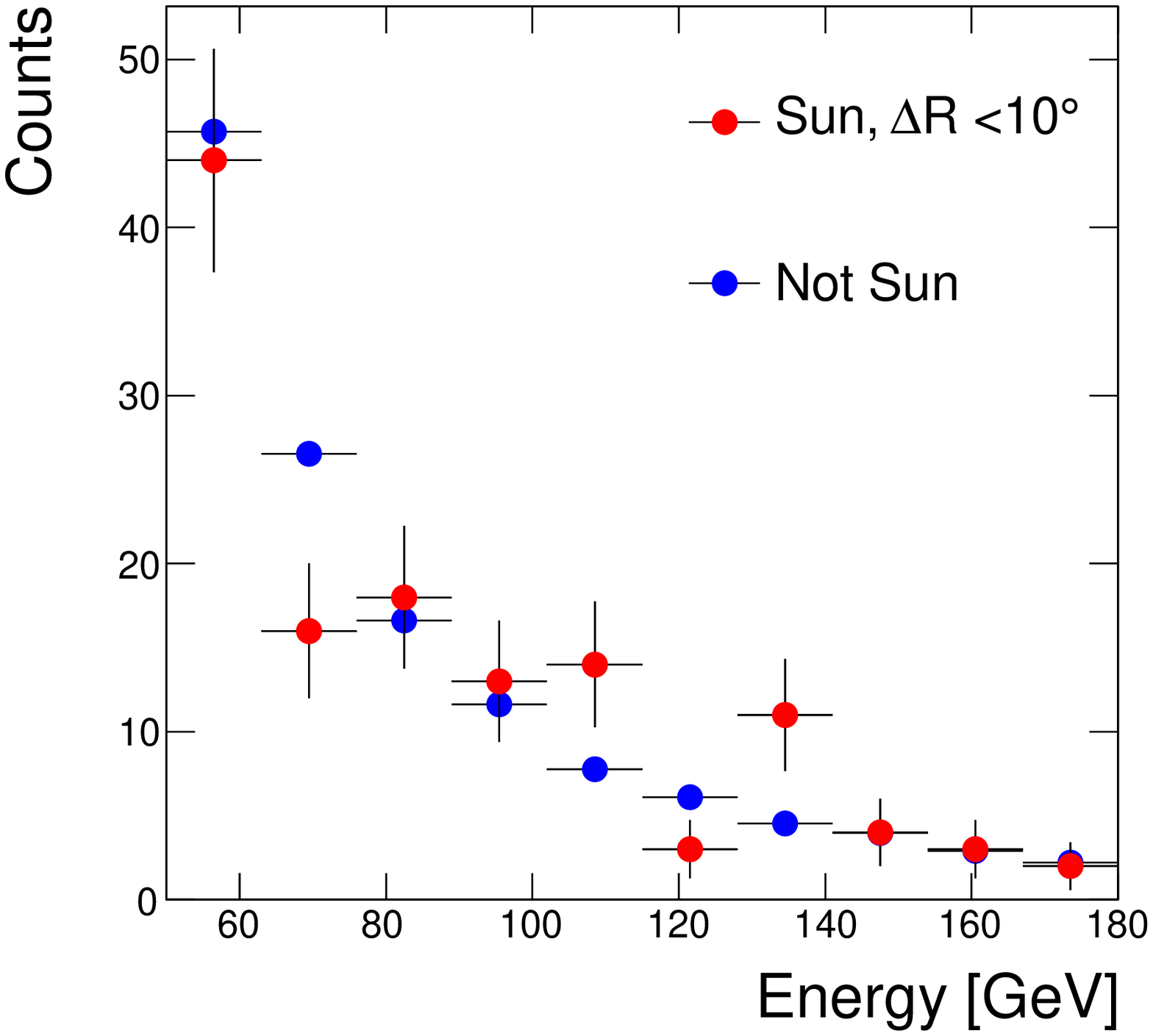}
\caption{ Clockwise from top left:  distribution of $\Delta
  R_{\astrosun}$ for photons with $E_\gamma \in [125,140]$ GeV; energy spectra for
  photons from the vicinity of the Sun using $\Delta
  R_{\astrosun}<3^\circ$, $\Delta R_{\astrosun}<10^\circ$, or $\Delta
  R_{\astrosun} \in[1,5]^\circ$, excluding the galactic center,
  compared to photons from the rest of the sky, again excluding the galactic
  center.}
\label{fig:sun2}
\end{figure}

\begin{figure}
\includegraphics[width=3in]{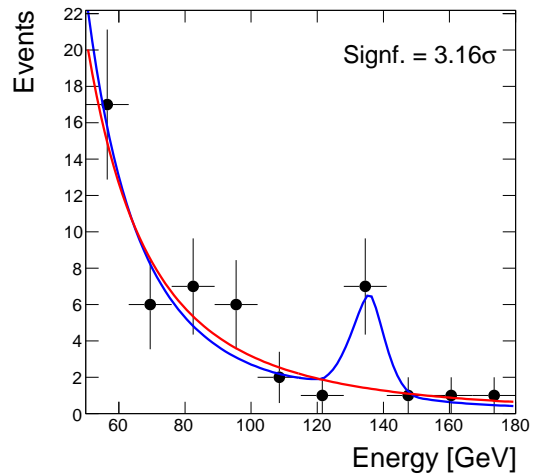}
\caption{Energy spectrum as shown in Fig.~\ref{fig:sun} for
  photons from the vicinity of the Sun ($\Delta
  R_{\astrosun}<5^\circ$), excluding the galactic center,
  with background-only and background-plus-signal hypotheses.}
  
\label{fig:signif_sun}
\end{figure}

\section{Instrumental Study with sPlots}

To consider potential instrumental explanations of the observed
features, we ask whether the photons in the peaks have distinct
instrumental characteristics from the background photons.  A naive approach
would be to simply divide the photons into two categories by energy
range; this leaves the peak-region sample with significant background
contamination.  A superior technique is statistical unfolding using
the {\sc sPlots} algorithm~\cite{splots}.

If a sample of events contains contributions from multiple sources,
whose relative contributions can be measured by fitting pdfs in a
discriminating variable, the  {\sc sPlots} algorithm can statistically extract
the distribution of each of the sources in other variables of
interest, which we refer to as the `unfolding variables'. Note that 
{\sc sPlots} requires knowledge of the pdf in the discriminating
variable, but no assumptions are required of the pdfs in the unfolding
variable other than that the pdfs can be factorized  between the discriminating and unfolding
variables. For a brief derivation of {\sc sPlots}, examples in toy
data and application to the Fermi-LAT galactic center spectrum, see Ref~\cite{fermisplots}.

\subsection{Analysis}

To analyze the features of the Fermi-LAT data using {\sc sPlots}, we must
fit the background and signal pdfs described above in the discriminating variable,
$E_\gamma$.  Applying these pdfs to the observed limb and Solar photon
energy spectrum yields the fits seen in Fig.~\ref{fig:fit1}.  In the
limb spectrum, we have not required $30^\circ < \theta < 45^\circ$ as
is done in Fig.~\ref{fig:signif}, but instead we apply a looser
requirement $\theta < 60^\circ$ in order to study the $\theta$-dependence.

\begin{figure}
\includegraphics[width=1.5in]{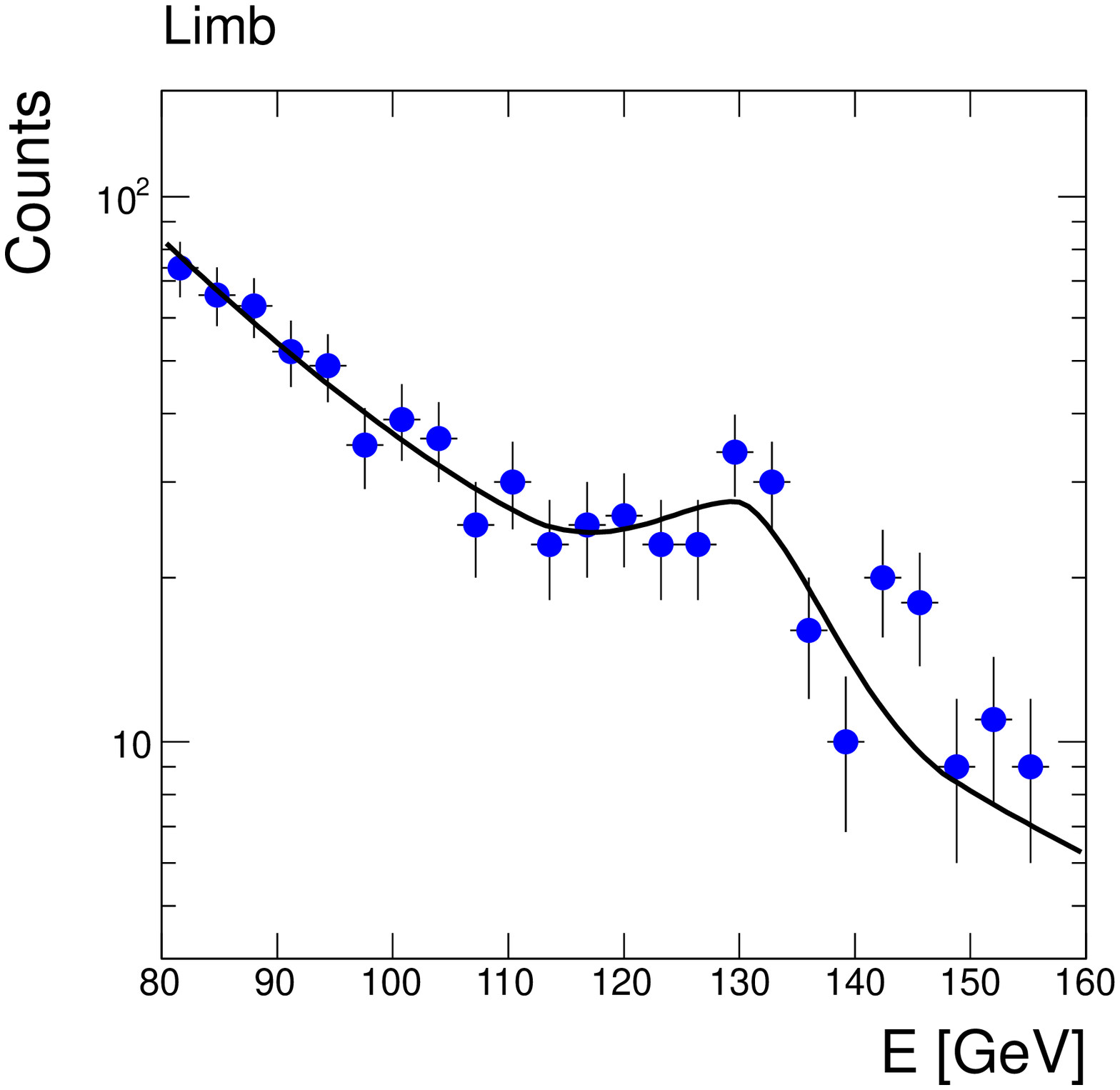}
\includegraphics[width=1.5in]{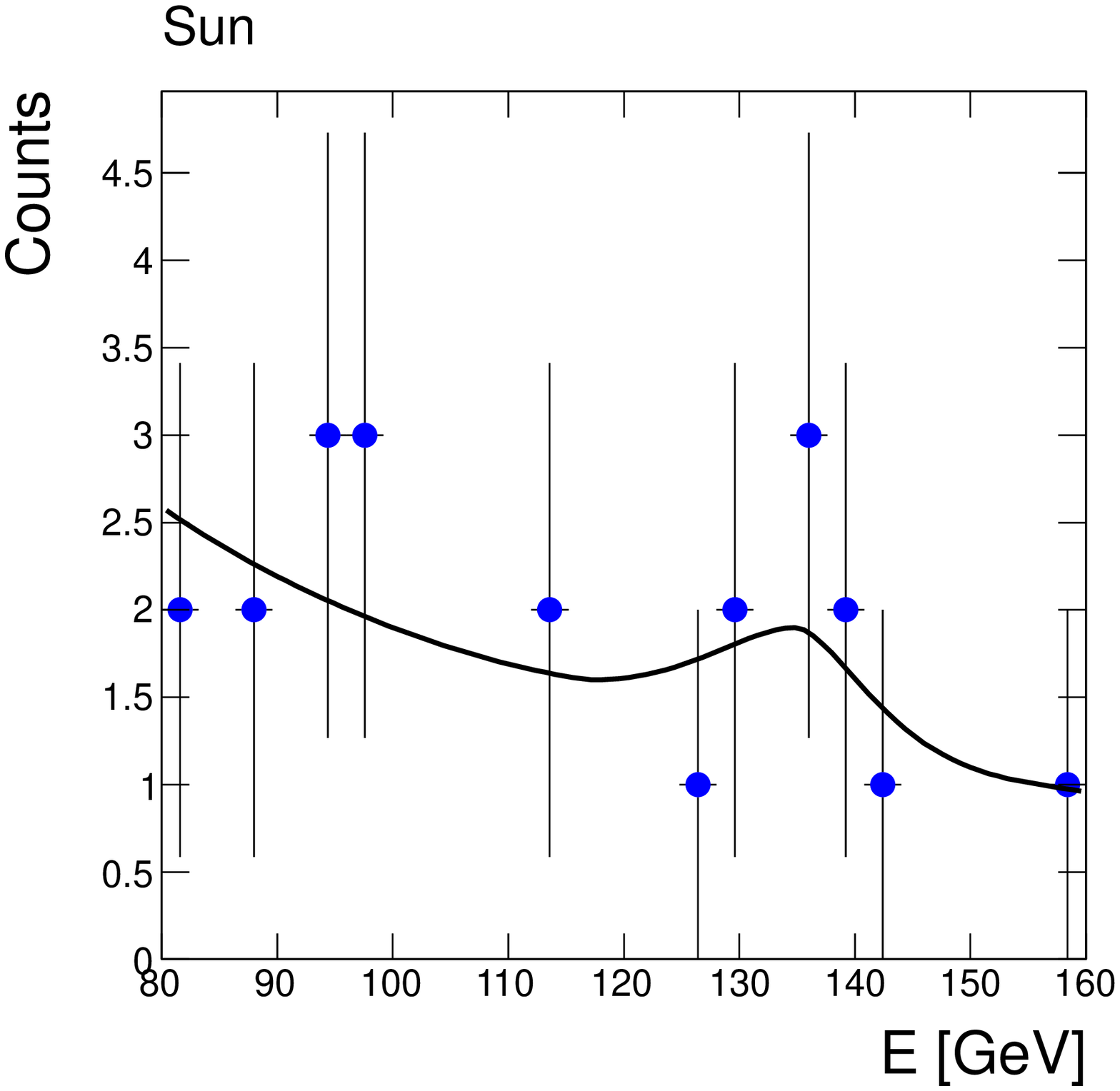}
\caption{Energy of Fermi-LAT photons with signal plus background
  fit. Left is the spectrum from the Earth's limb with $\theta < 60^\circ$,
  right is the Solar spectrum with $\Delta
  R_{\astrosun}<5^\circ$, excluding the galactic center.}
\label{fig:fit1}
\end{figure}

Unfolded distributions of incidence angles are shown in
Fig.~\ref{fig:detang1}. The recorded time and conversion type are
in Fig.~\ref{fig:timeback1}. The magnetic field parameters are shown
in Fig.~\ref{fig:mag1}.

 In each case, we compare the
distributions quantitatively by calculating the $\chi^2/$dof between
the peak and background distributions. As the signal and background
weights are anti-correlated, this is calculated as

\[ \chi^2 =
\sum_{\textrm{bin}\ i}\frac{(N^i_{\textrm{peak}}-N^i_{\textrm{bg.}})^2}{(\Delta N^i_{\textrm{peak}}+\Delta
  N^i_{\textrm{bg.}})^2}\]

\noindent
where $N^i_{\textrm{peak}}$ is the sum of the weights
$\sum sP_{\textrm{peak}}$ in that bin, and $\Delta
N^i_{\textrm{peak}}$ is calculated from toy simulations which estimate
the expected variance of the measurement in each bin.  Similar expressions apply for the
background uncertainties.

\begin{figure}
\includegraphics[width=1.6in]{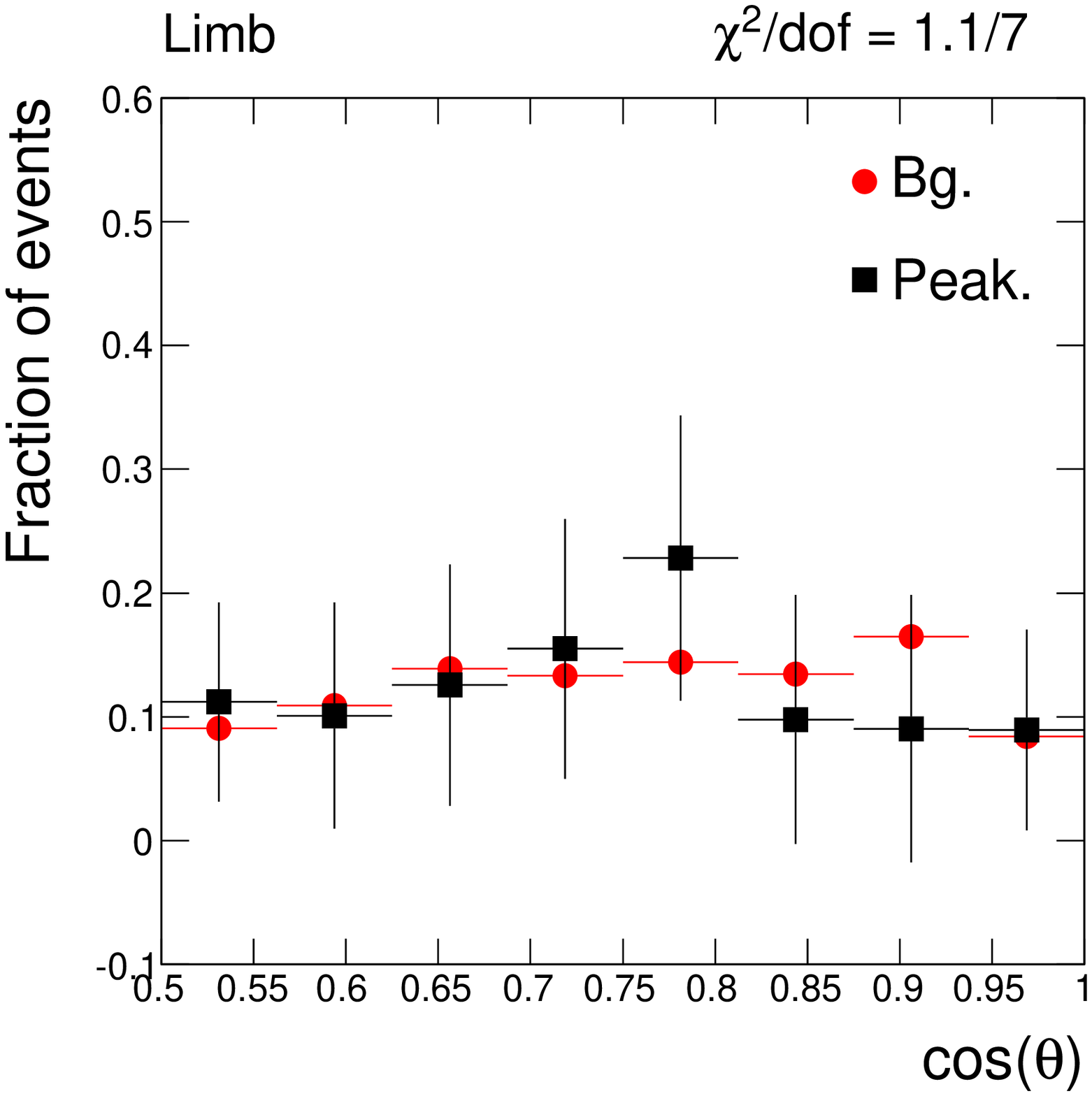}
\includegraphics[width=1.6in]{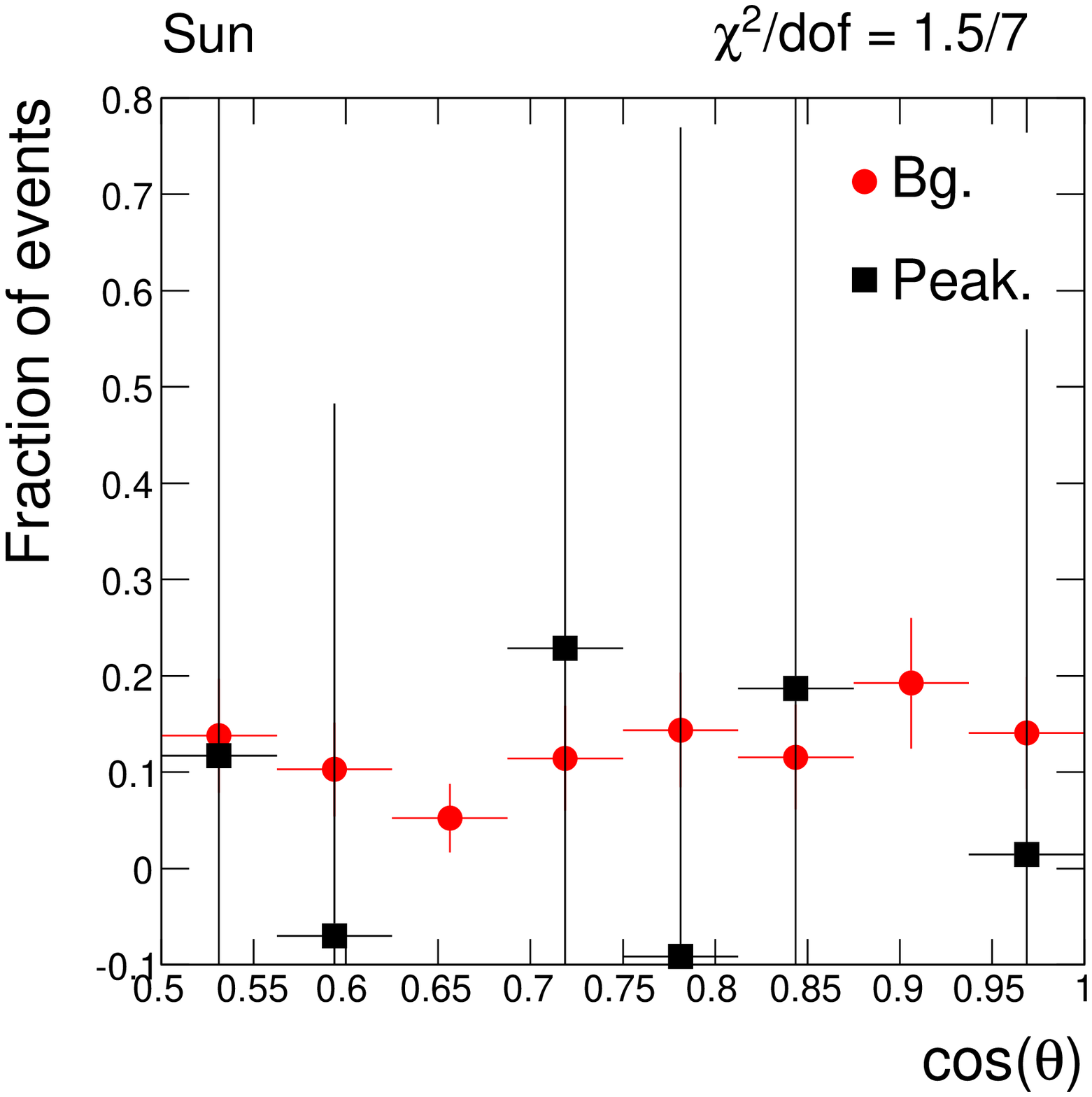}\\
\includegraphics[width=1.6in]{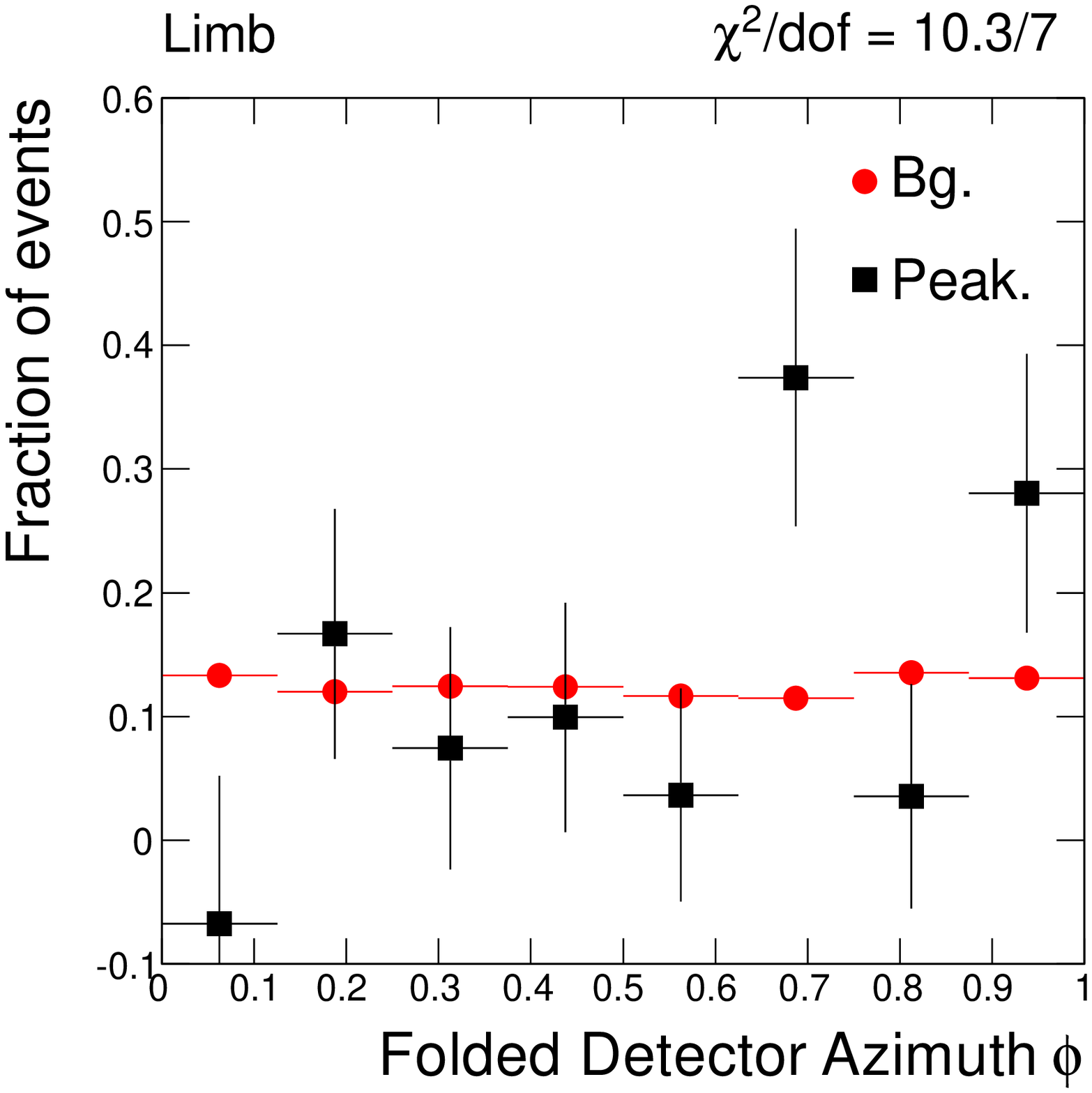}
\includegraphics[width=1.6in]{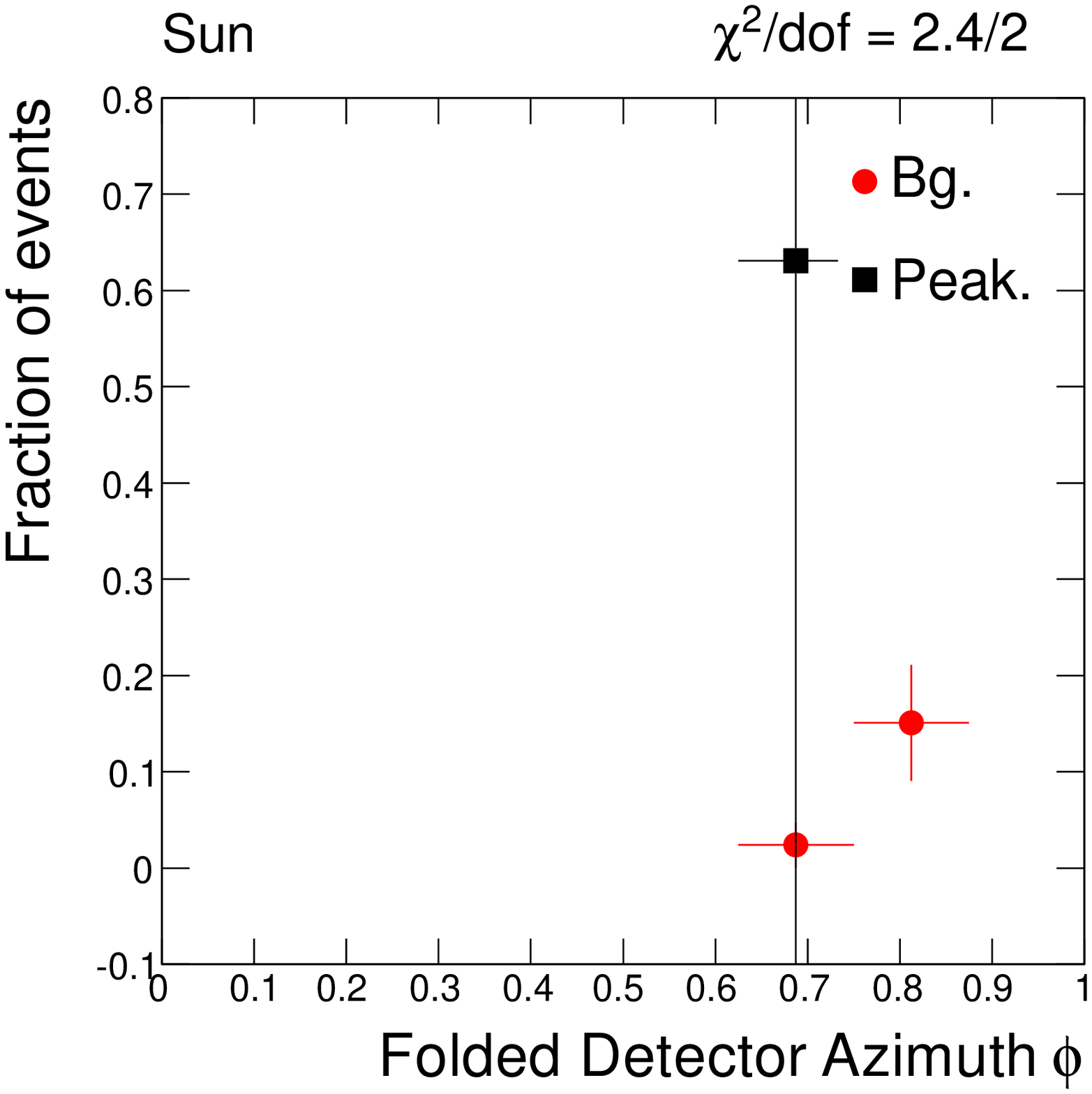}
\caption{Disentangled signal and background distributions. Top,  $\cos(\theta)$ where
  $\theta$ is the photon incidence angle relative  to a line normal
  the Fermi-LAT face. Bottom, $\phi$, the photon incidence angle
  relative to the sun-facing side~\cite{fermidefs}. Left are photons from the
  Earth's limb, right are photons from the vicinity of the Sun. For the Solar
  spectrum,  $\phi$ is aligned with respect to Fermi's solar panels,
  which track the Sun.}
\label{fig:detang1}
\end{figure}

\begin{figure}
\includegraphics[width=1.6in]{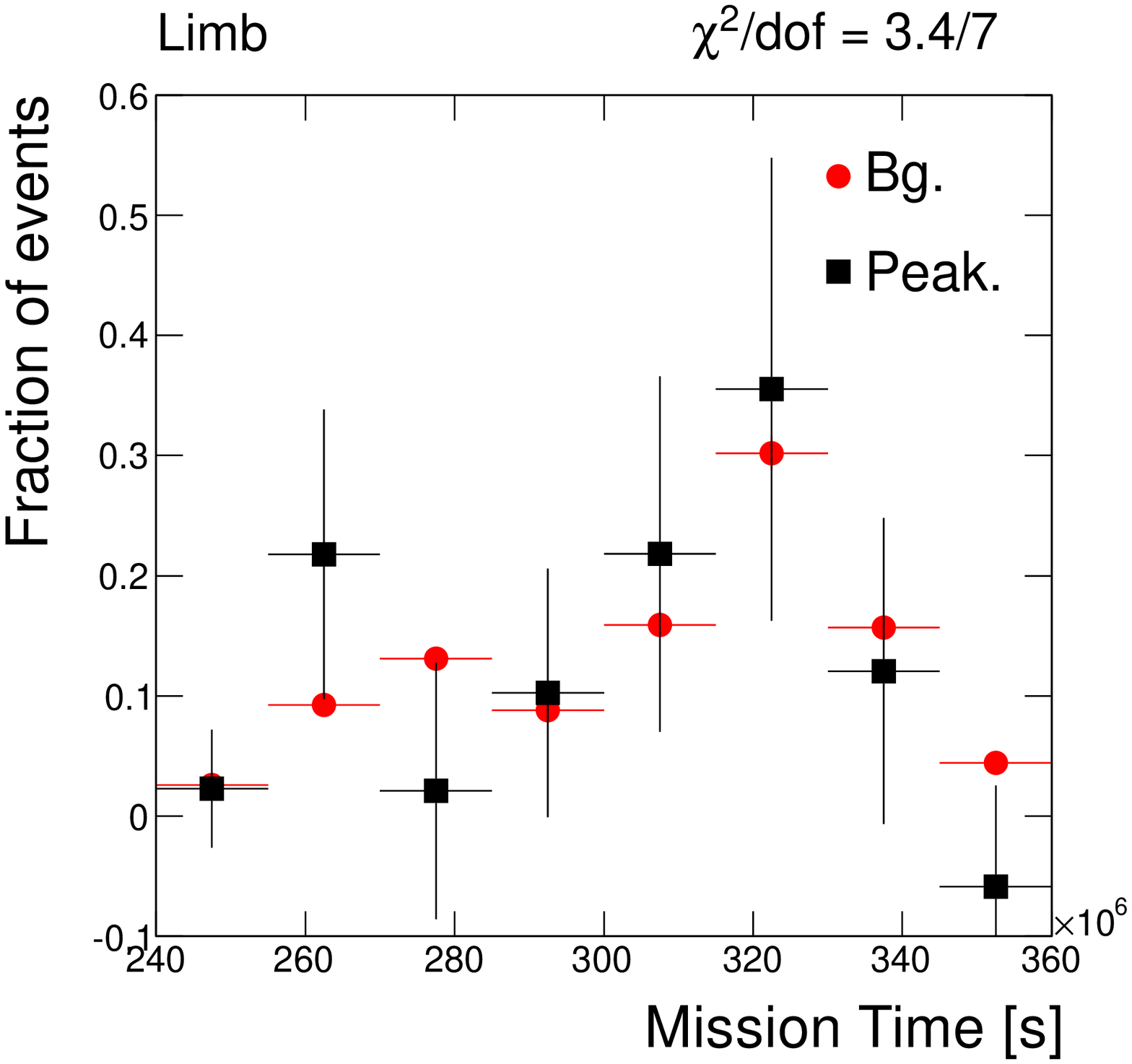}
\includegraphics[width=1.6in]{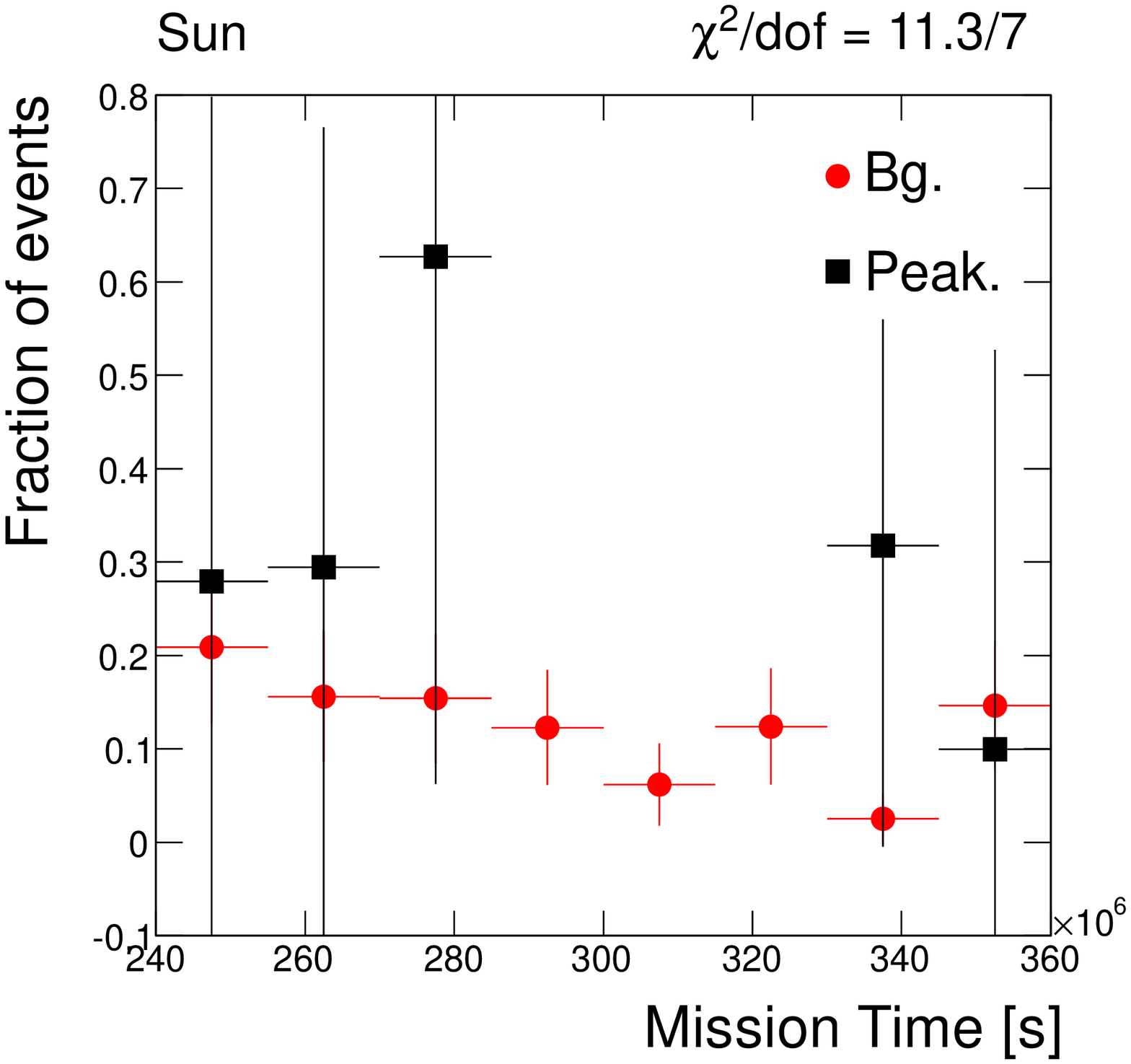}
\includegraphics[width=1.6in]{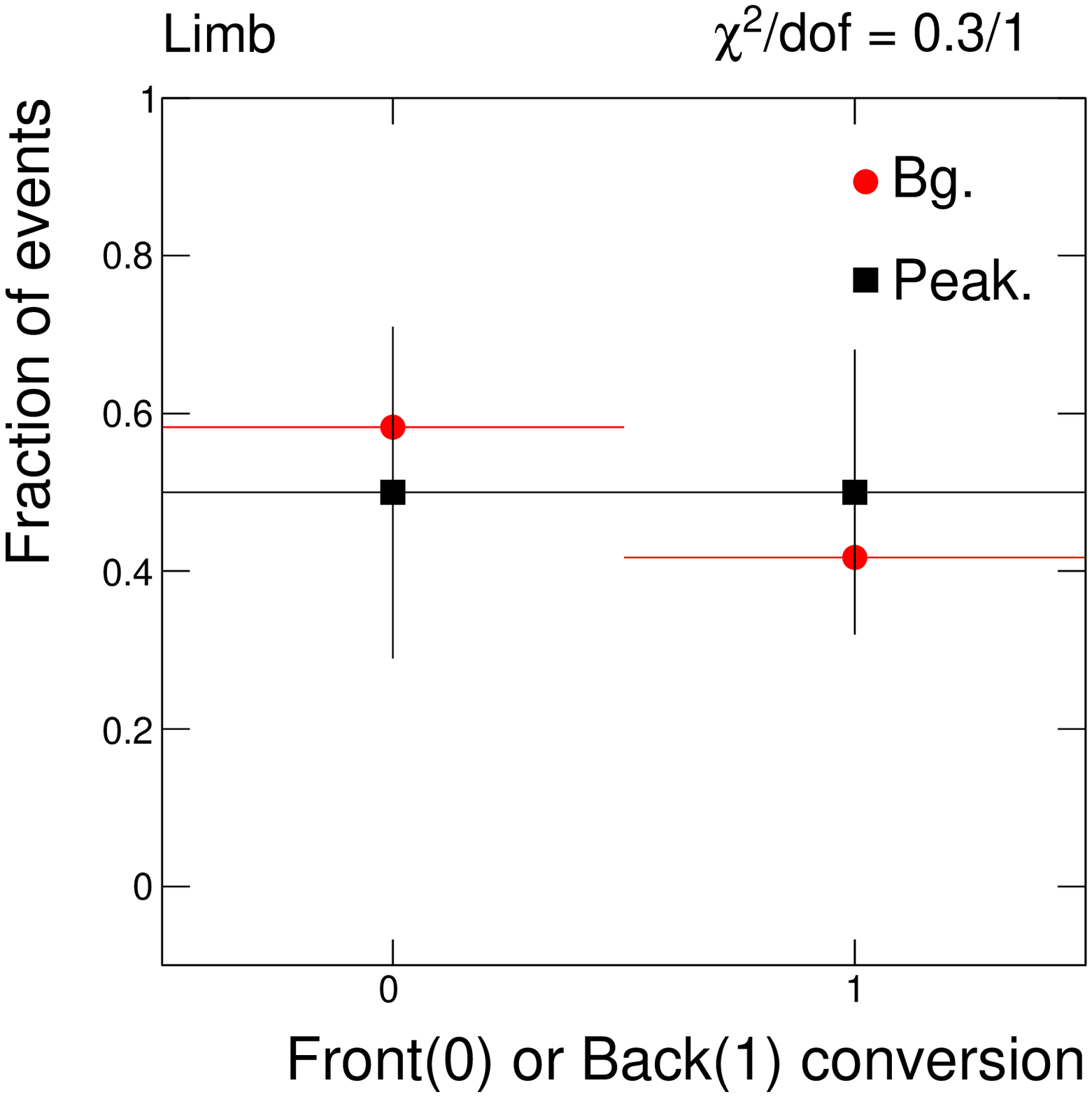}
\includegraphics[width=1.6in]{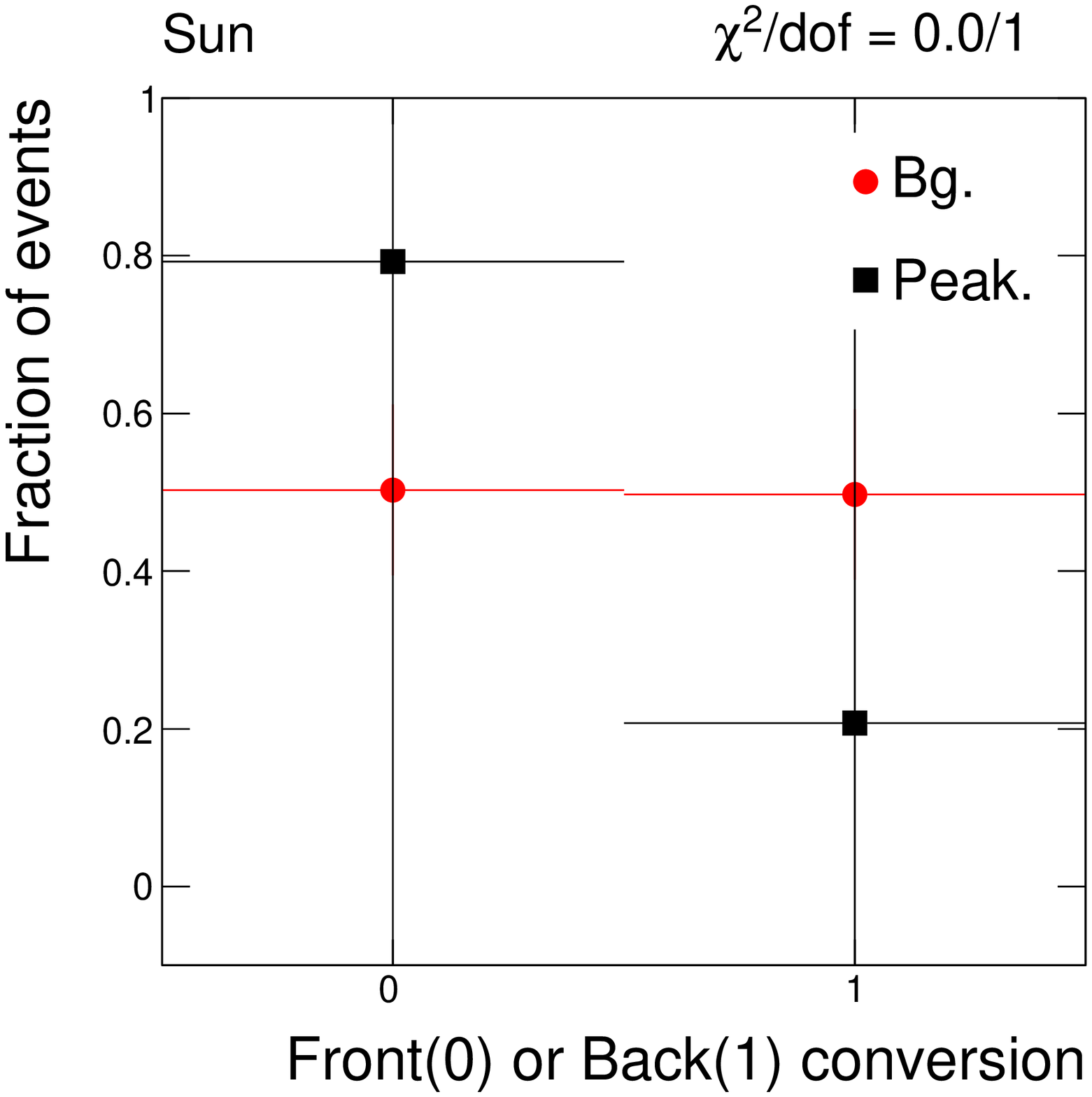}
\caption{Disentangled signal and background distributions.  Top, the mission elapsed time since Jan
  1 2001~\cite{fermidefs}. Bottom, 
  fraction of events in which the pair production is induced in the
  front (thin) or back (thick) layers of the tracker.  Left are photons from the
  Earth's limb, right are photons from the vicinity of the Sun.}
\label{fig:timeback1}
\end{figure}

\begin{figure}
\includegraphics[width=1.6in]{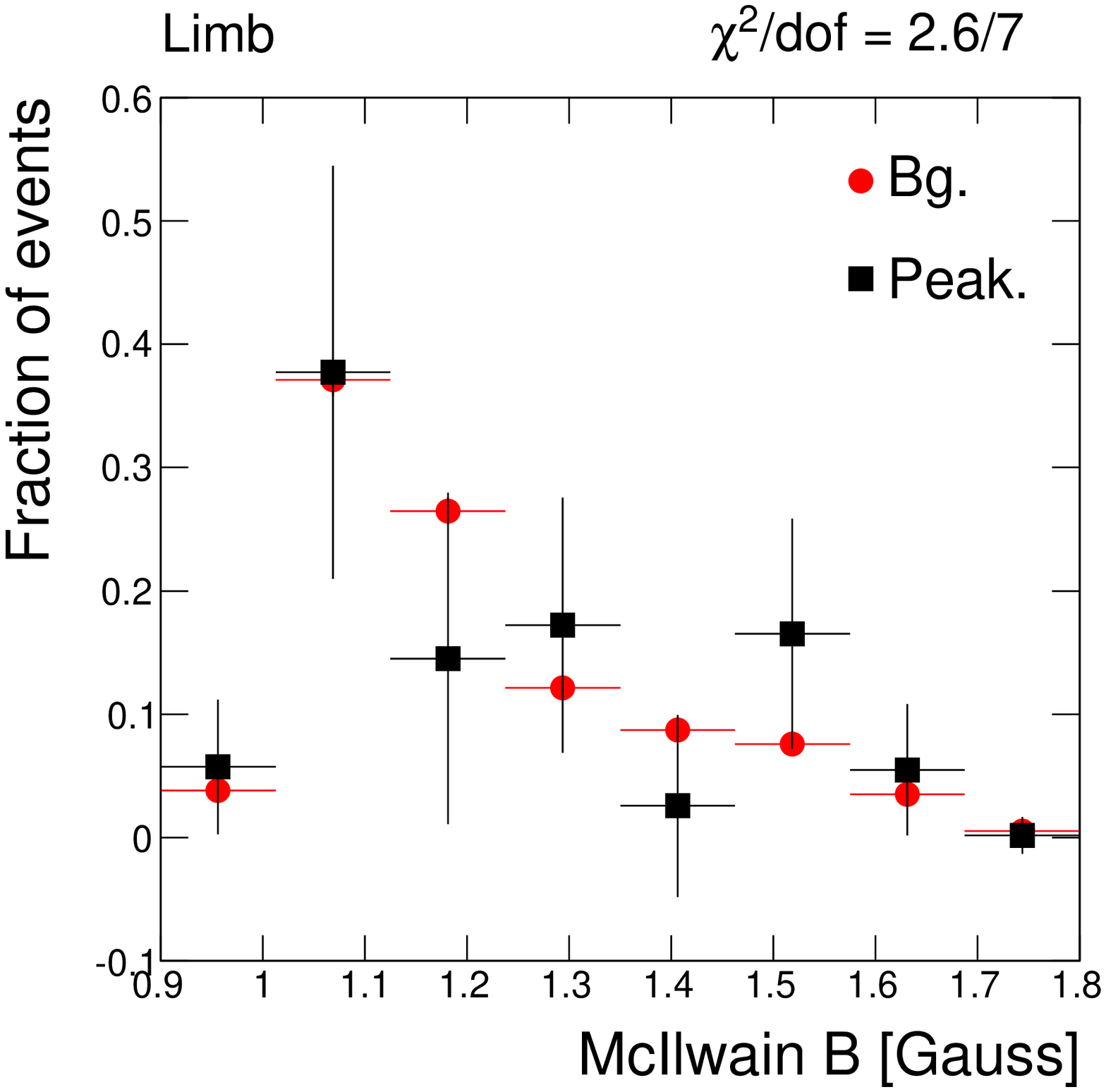}
\includegraphics[width=1.6in]{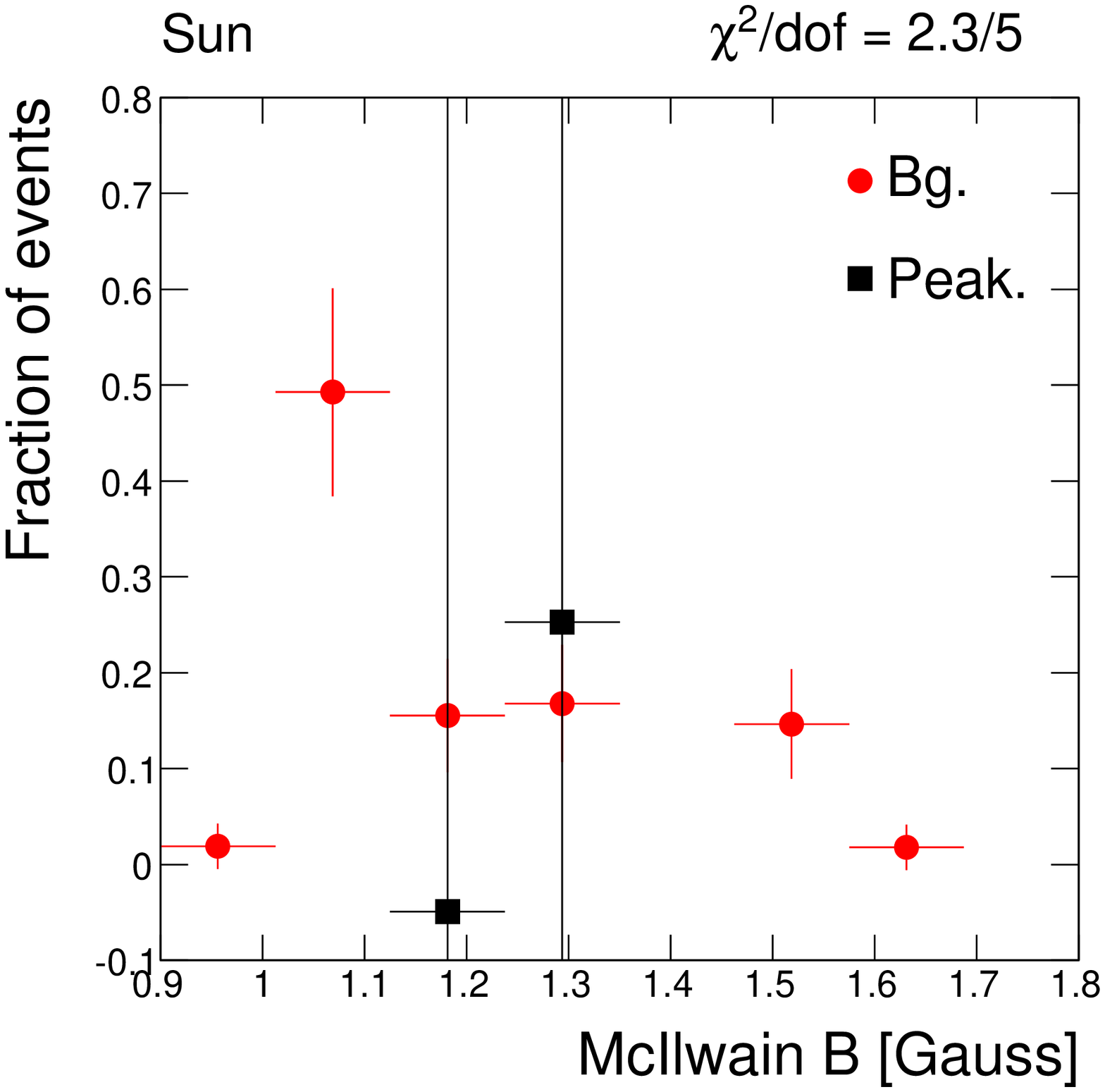}
\includegraphics[width=1.6in]{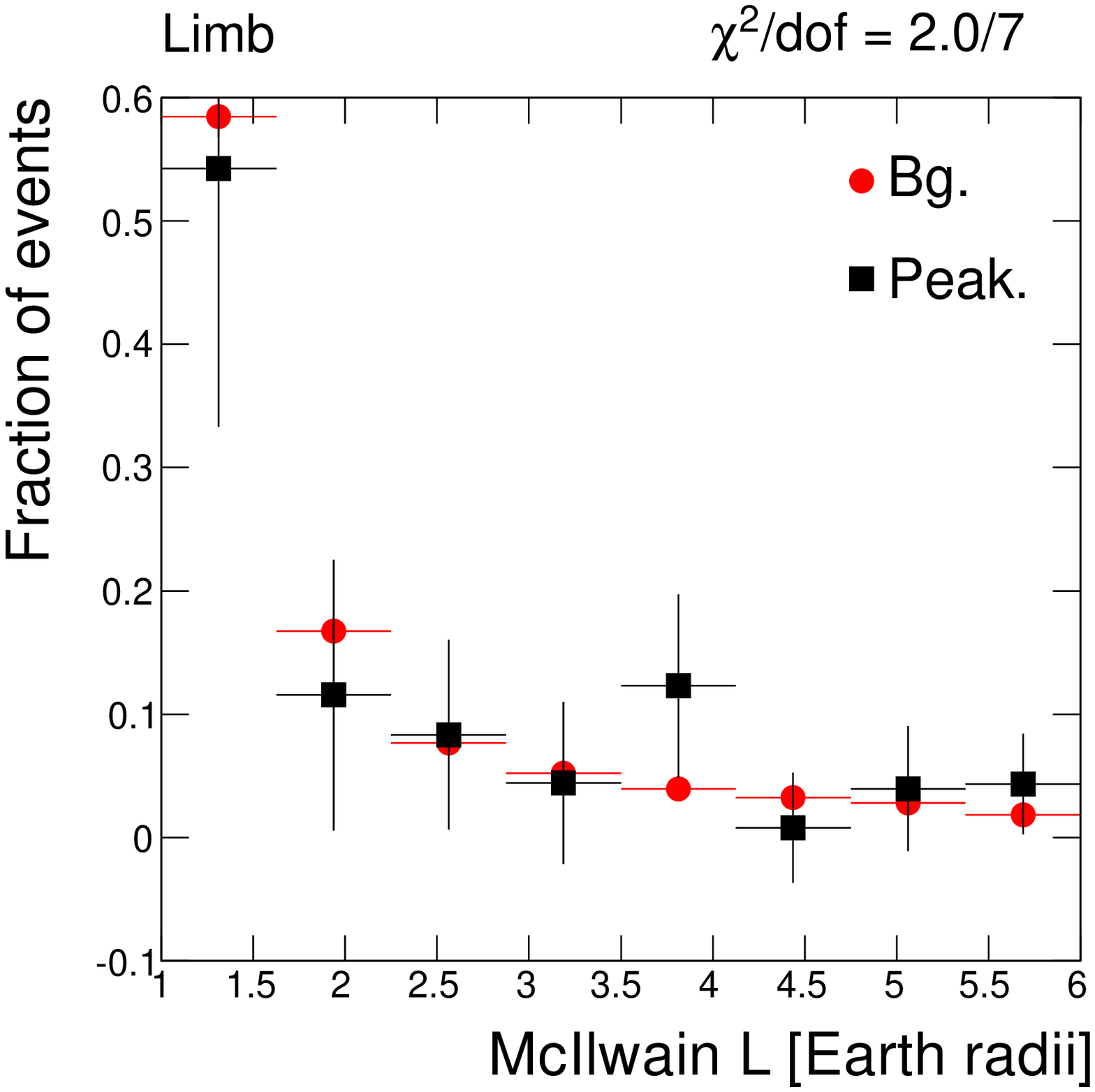}
\includegraphics[width=1.6in]{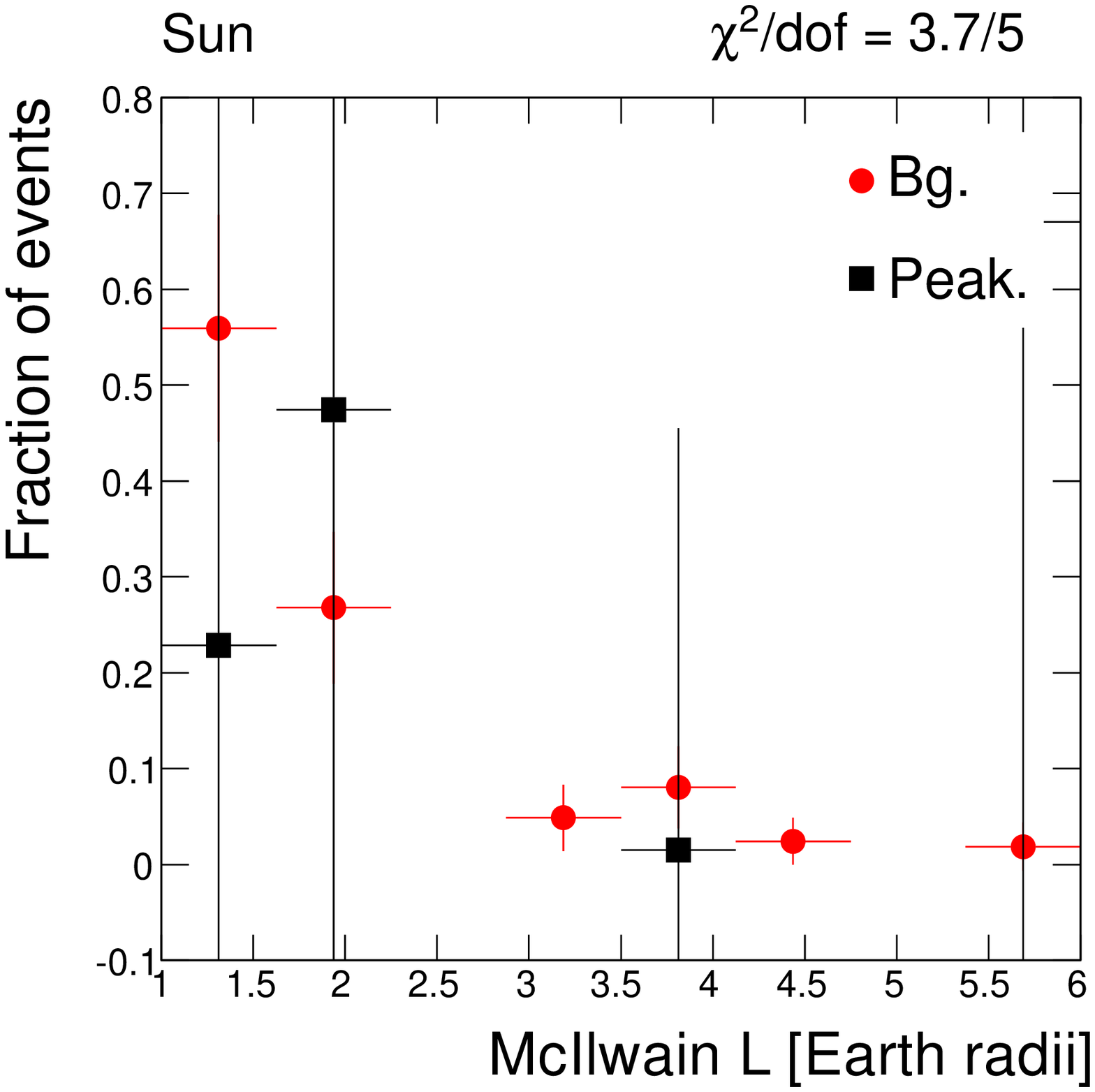}
\caption{Disentangled signal and background distributions.  Top,
  magnetic field strength in terms of the McIlwain $B$
  parameter. Bottom, the McIlwain $L$ parameter.  Left are photons from the
  Earth's limb, right are photons from the vicinity of the Sun.}
\label{fig:mag1}
\end{figure}

\subsection{Discussion}

Our previous analysis of the galactic center feature showed that a
large fraction of peak photons had $\cos(\theta)$ near 0.7, unlike the
photons from the background (Fig. 5 in Ref.~\cite{fermisplots}).  The limb feature photons also peak near
this region, first reported in Ref~\cite{finksu}, and confirmed here,
see Fig.~\ref{fig:detang1}.  The Solar spectrum statistics are too poor
to contribute to this question.

The galactic center instrumental analysis also showed some discrepancy
in the magnetic field environement of the LAT when the photons were
recorded (Fig. 11 in Ref.~\cite{fermisplots}), near 1.6 Gauss.  The
limb photons have a minor discrepancy near 1.5 Gauss, but entirely
consistent with statistical fluctuations.  Similarly, the galactic
center photons have a feature near folded azimuth $\phi=0.8$; the limb photons
have some features near $\phi=0.7$ and $\phi=0.9$, but a deficit near
$\phi=0.8$ with respect to the background photons.

\section{The rest of the sky}

If the feature near $E_\gamma=130$ GeV has an instrumental rather than
astrophysical explanation, then it should be independent of the source
of the photons and one should be able to identify similar features in
any spectrum.  The full-sky spectrum with no restrictions on
instrumental characteristics does not show such a feature; if it
exists, a feature in the full-sky spectrum may be localized to
particular instrumental regions and washed out in the unrestricted spectrum.

  To discover potential
instrumental causes, we look for universal characteristics of the
peaks found in the three spectra.  If the full sky spectrum in
a restricted range of these instrumental characteristics were to reveal a feature
near $E_\gamma=130$ GeV, it would lend support to the instrumental explanation.

We first consider the incident angle, $\theta$, already shown to reveal
the Earth's limb peak.  In Fig.~\ref{fig:quad_theta}, we show the
spectra in the three regions defined earlier: the galactic center, the
Sun, and the Earth's limb. In addition, we show the spectrum in the
rest of the sky, defined by requiring zenith angle $<105^\circ$, rocking angle
  $<52^\circ$, $\Delta R_{\astrosun} >5^\circ$, and
  $\sqrt{b^2+l^2}>5^\circ$. These four regions are disjoint.

First, we examine the spectrum with no restrictions on $\theta$. There
are features in the galactic center and Solar spectrum, but nothing
evident in the limb or sky.  When we restrict $\theta$ to $[30,45]^\circ$,
the limb feature is evident, and we appear to capture a large 
fraction of the galactic center and Solar features.  Most
interestingly, a small feature is evident in the sky spectrum near
$E_\gamma=130$ GeV. This
feature is intriguing, but is not large enough to unambiguously confirm an instrumental
issue in the region $\theta$ to $[30,45]^\circ$.

A similar analysis using the detector azimuth (Fig.~\ref{fig:quad_phi}) is able to capture most
of the galactic center and Solar spectra, but do not reveal features
in the limb or sky spectra.

\begin{figure*}
\includegraphics[width=\textwidth]{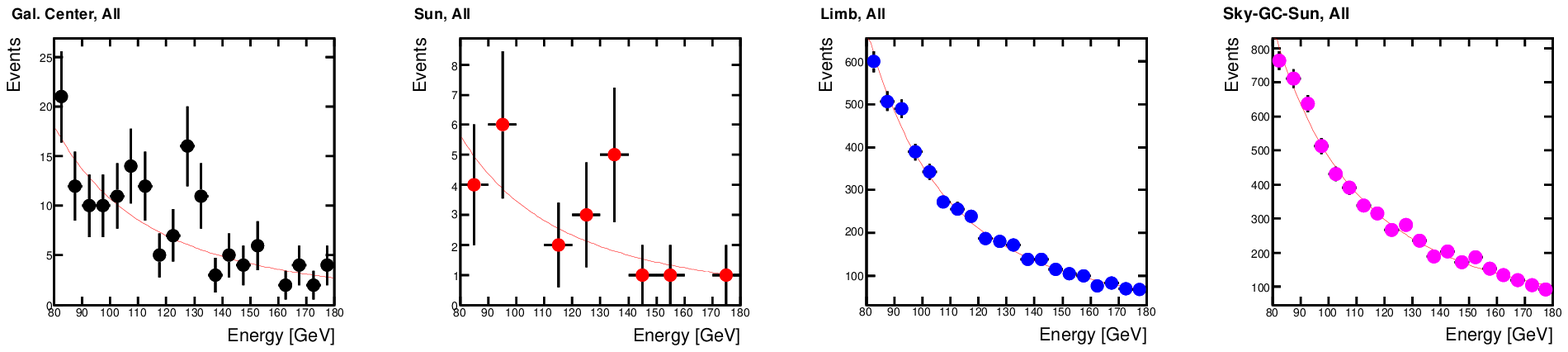}
\includegraphics[width=\textwidth]{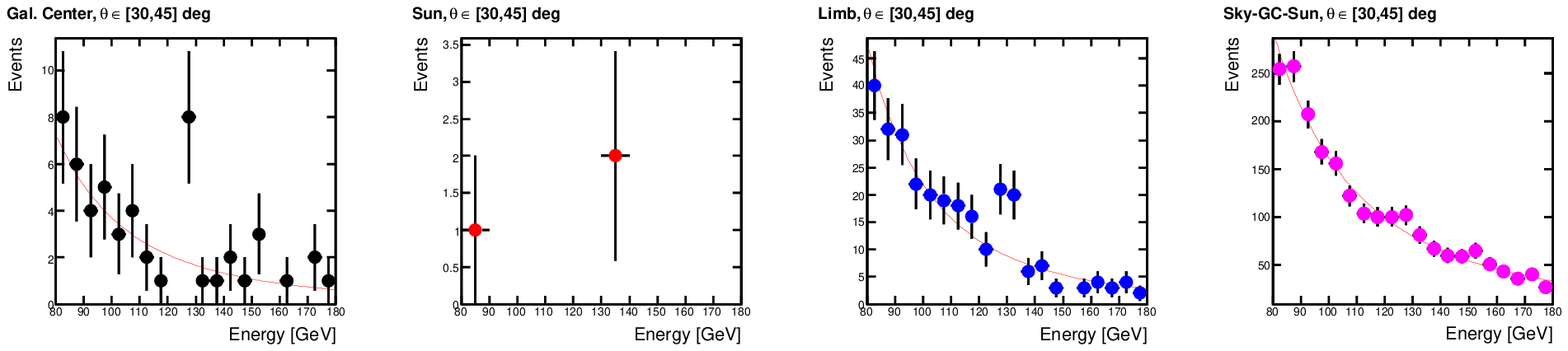}
\includegraphics[width=\textwidth]{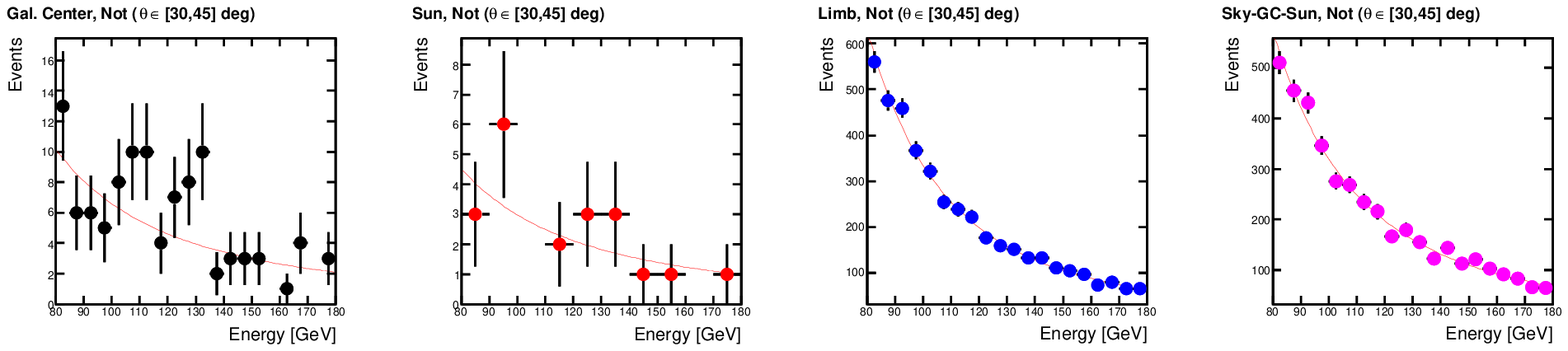}
\caption{ Energy spectra for four regions; from left: galactic center,
  Solar vicinity, Earth's limb, and the full sky with the GC and Sun
  removed. Top shows the complete spectrum, center is for photons with
  incident angle $\theta$ in the range $[30,45]$ degrees, and bottom
  for photons with $\theta$ out of the range $[30,45]$ degrees.}
\label{fig:quad_theta}
\end{figure*}

\begin{figure*}
\includegraphics[width=\textwidth]{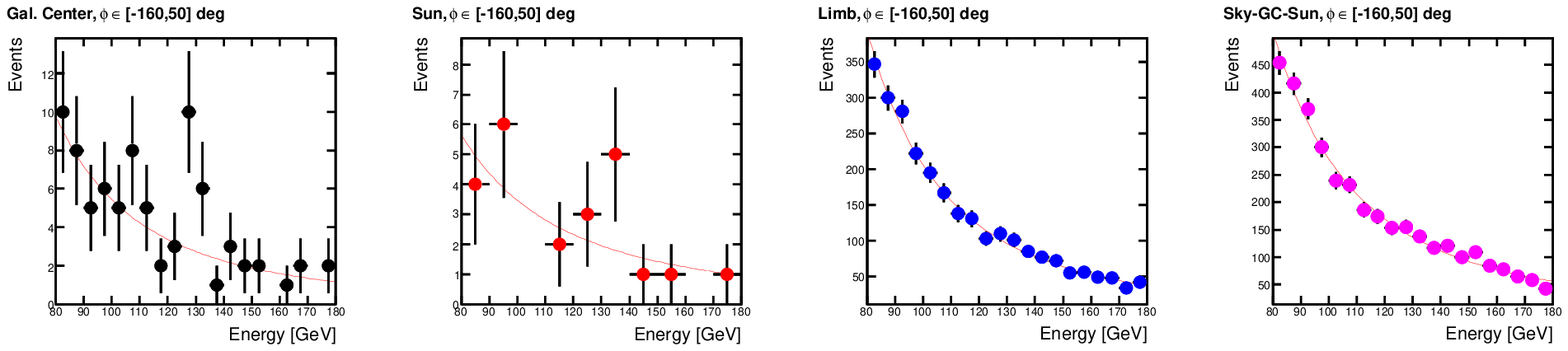}
\includegraphics[width=\textwidth]{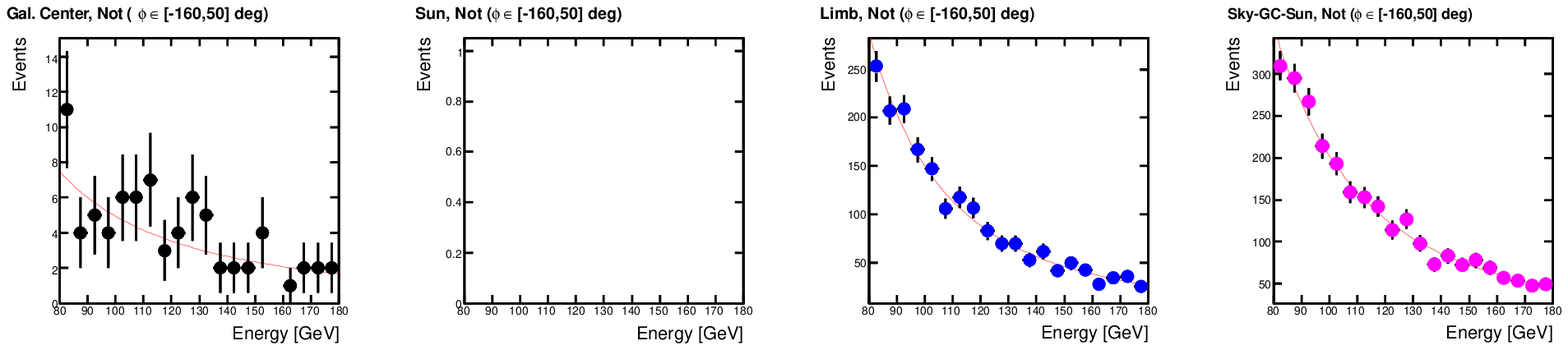}
\caption{Energy spectra for four regions; from left: galactic center,
  Solar vicinity, Earth's limb, and the full sky with the GC and Sun
  removed. Top is for photons with
  azimuth angle $\phi$ in the range $[-160,50]$ degrees, and bottom
  for photons with $\phi$ out of the range $[-160,50]$ degrees.}
\label{fig:quad_phi}
\end{figure*}

\section{Conclusions}

We report a spectral feature near $E_\gamma=130$ GeV in the spectrum
of photons from the vicinity of the Sun, where none is expected from
dark matter annhiliations.  The statistical significance is
$3.2\sigma$, corresponding to a $p$-value of $<0.5$\% that the feature
is a statistical fluctuation. This is the first report of such a
feature. Based on current understanding of the
Solar dark matter halo, if the feature in the Solar spectrum is
more than statistical fluctuation, it represents evidence of instrumental
issues, not dark matter annihiliation. 

We also analyze the instrumental characteristics of the Earth's limb
feature, and attempt to identify a universal characteristic of photons
in the three observed features using the {\sc sPlots} algorithm.  

Finally, we show that a narrow
range of incident photon angles includes many photons from the
observed peaks, and reveals a feature near $E_\gamma=130$ GeV in the
spectrum of the remainder of the sky. While we attempt to
remove dependence on photon source by examining the full sky and
placing restrictions only on $\theta$,  instrumental characteristics
may in general be
correlated with the photon source, as the geometry of the Sun, Earth
and galactic center relative to the LAT may give different average
distributions of instrumental characteristics.  If the spectral
feature were caused by an instrumental issue in a slice of $\theta$
and another not-yet-identified variable, it is plausible that photons
from the Earth, Sun and galactic center sweep across this slice
differently, giving different $\theta$-dependences and explaining the stronger
$\theta$-dependence of the limb and sky features than in the solar and
galactic center features. In addition, more precisely
focusing on photons in such a slice would enhance the feature in the sky spectrum.

 While this does not give a
definitive instrumental explanation for these spectral features, it
casts significant further doubts on the dark matter hypothesis.

\acknowledgments
\section{Acknowledgements}

DW acknowledges  contributions, explanations
and useful discussions with Eric Albin which are clearly deserving of
authorship, and comments from Itay Yavin, Andrew Nelson, and Kanishka Rao.
DW is supported by grants from the Department of Energy
Office of Science and by the Alfred P. Sloan Foundation. DW is
grateful to the Aspen Center for Physics, where this some of this work was
performed and supported by NSF grant no. 1066293.

\end{document}